\documentstyle[12pt,epsf]{article}
\begin{document}

\newcommand{\be}{\begin{equation}}
\newcommand{\ee}{\end{equation}}
\newcommand{\nn}{\nonumber}
\newcommand{\rpv}{\mbox{$\rlap{\kern0.25em/}R_p$}}
\newcommand{\rp}{\mbox{$R_p$}}
\renewcommand{\l}{\lambda}

\newcommand{\met}{\mbox{$\rlap{\kern0.25em/}E_T$}}
\renewcommand{\topfraction}{1}
\renewcommand{\bottomfraction}{1}

\begin{titlepage}
\begin{flushright}
      hep-ph/9704343     \\
       April 1997        \\
\end{flushright}
\vspace{2cm}
\begin{center}
{\Large  \bf
Could we learn more about HERA high $Q^2$ anomaly from
    LEP200 and TEVATRON ? }

\vspace{1cm}
{\bf A.S.Belyaev} $^{\mbox{a,}}$\footnote{e-mail:
belyaev@monet.npi.msu.su},
{\bf A.V.Gladyshev} $^{\mbox{b,}}$\footnote{e-mail:
gladysh@thsun1.jinr.dubna.su}

\vspace{1cm}

$^{\mbox{a}}$ {\it Skobeltsin Institute for Nuclear Physics,
Moscow State University, \\ 119 899, Moscow, RUSSIA}

\vspace{.5cm}

$^{\mbox{b}}$ {\it Bogoliubov Laboratory of Theoretical Physics,
     Joint Institute for Nuclear Research, \\
	141 980 Dubna, Moscow Region, RUSSIA}

\end{center}

\vspace{1cm}

\begin{abstract}
The excess of high $Q^2$ events at HERA  has been the subject of recent
extensive studies in the framework of several models related to new
physics.  Here, we would like to concentrate on the most promising,
from our point of view, model describing HERA anomaly.  We investigate
HERA events within the $R$-parity broken SUSY model and check its
relation to LEP and TEVATRON colliders. This study shows that if a
squark  resonance really takes place at HERA, supersymmetry with broken
$R$-parity can be  revealed at either LEP200 or TEVATRON.
\end{abstract}

\end{titlepage}


\section{ Introduction}
Anomalous high-$Q^2$ events recently observed at HERA by the
H1~\cite{h1} and ZEUS~\cite{zeus} collaborations have provoked much
efforts of their explanation within different theoretical
frameworks~\cite{many,ellis}.  The proposed explanations are contact
interactions, leptoquarks and signals of $R$-parity violation  in
supersymmetric interactions. The contact interactions do not give  us
satisfactory explanation  of either the shape of $Q^2$ distribution of
HERA events or the distribution over the $e+jet$ invariant mass which
has a peak around 200 GeV. We would like to take a fresh look at the
possibility of $R$-parity violation providing a $s$-channel resonance at
HERA.

The starting point for our analysis is the superpotential of the
Minimal Supersymmetric Standard Model with $R$-violating
terms included
\begin{eqnarray}
W &=&  h_E L E^c H_1 + h_D Q D^c H_1 + h_U Q U^c H_2 +
\mu H_1 H_2  \nn \\
&& + \mu_i H_2 L_i+
\l_{ijk} L_i L_j E^c_k +
\l'_{ijk} L_i Q_j D^c_k +
\l''_{ijk} U^c_i D^c_j D^c_k
\end{eqnarray}

The terms with $\l$ and $\l'$ violate a lepton number while the last
term with $\l''$ violates a barion number. $B$ and $L$ operators, if
they exist simultaneously, lead to a fast proton decay~\cite{e37}.
Thus, only one kind of operators is allowed. HERA events can be
explained in the framework of the $R$-parity broken SUSY model only if
$\l'\ne 0$. That is why we hereafter consider the case when $\l''=0$.

The  $\l$ and $\l'$ -terms result in the following lagrangian for
component fields:

\begin{eqnarray}
L_{\rpv,\l}&=&\l_{ijk}[\tilde{\nu}_{iL}\bar{e}_{kR}e_{jL}+
\tilde{e}_{jL}\bar{e}_{kR}\nu_{iL}+
{\tilde{e}^*}_{kL}\overline{(\nu_{iL})^C}e_{jL} -
(i\leftrightarrow j)]_F + h.c.,\\
L_{\rpv,\l'}&=&\l'_{ijk}[\tilde{\nu}_{iL}\bar{d}_{kR}d_{jL}+
\tilde{d}_{jL}\bar{d}_{kR}\nu_{iL}+
{\tilde{d}^*}_{kL}\overline{(\nu_{iL})^C}d_{jL} - \nn \\
&&\tilde{e}_{iL}\bar{d}_{kR}u_{jL}-\tilde{u}_{jL}\bar{d}_{kR}e_{iL}-
{\tilde{d}^*}_{kL}\overline{(e_{iL})^C}u_{jL}]
 + h.c.
\end{eqnarray}

The present experimental limits on different couplings
are presented in table~\ref{limits}.

\begin{table}[thb]
\begin{tabular}{p{2.1cm}p{0.8cm}|p{2.5cm}p{0.8cm}|p{2.8cm}p{0.8cm}|p{2.1cm}p{0.8cm}}
  $\l_{13k}\le 0.2                $ & (a)\cite{e35}
& $\l'_{111}\le 7\cdot{10^{-3}}$ & (c)\cite{e20}
& $\l'_{131}\le 3.7\cdot {10^{-2}}$ & (b)\cite{e37}
& $\l'_{23k}\le 0.24           $ & (a)\cite{e35} \\
  $\l_{23k}\le 0.2                $ & (a)\cite{e35}
& $\l'_{ijk}\le 0.024          $ & (d)\cite{e21}
& $\l'_{132}\le 7.8\cdot {10^{-3}}$ & (b)\cite{e37}
& $\l'_{ijk}\le 0.52$            & (e)\cite{b17} \\
  $\l_{12k}\le 0.08               $ & (a)\cite{e35}
&                     &
& $\l'_{133}\le 1.7\cdot {10^{-3}}$ & (b)\cite{e37}
&                     &
\end{tabular}
\caption{ \label{limits} 
The strictest experimental bounds on $R$-parity violating
couplings.
Limits derived from the following processes:
a) charged-current universality,
b) $\nu_e$ mass;
c) neutrinoless double beta-decay;
d) $BR(K^+\to\pi^+\nu\bar{\nu})$;
e) atomic parity violation and  $eD$ asymmetry.}
\end{table}

Bounds shown in table~\ref{limits}  scale with the following factors:
\par
a), d) and e) --  $M_{\tilde{q}} / 200 \ GeV$;
\par
b) --             $\sqrt{M_{\tilde{q}} / 200 \ GeV}$;
\par
c) -- $(M_{\tilde{q}}/200\ GeV)^2\sqrt{M_{\tilde{g}}/1\ TeV}$

The purpose of the present letter is to re-examine the possible
explanation of the excess of high $Q^2$ neutral current DIS events at
HERA assuming $R$-parity violation and discuss its possible
check at LEP200 and TEVATRON.

It should be stressed that  all analytical and the most part of numerical
calculations have been made by the CompHEP package~\cite{comphep}.This package
allows one to perform the complete tree level calculations in the framework of
any fed model. We have implemented into this program the part of supersymmetric
standard model with R-parity violating terms  relevant for our analysis.

\begin{figure}[thb]
  \vspace*{-4.0cm}
  \hspace*{1.0cm}
 \begin{center}
    \leavevmode
    \epsfxsize=15cm
    \epsffile{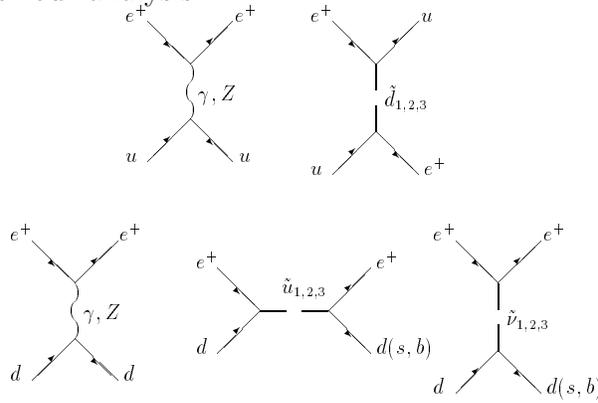}
    \vspace*{-11.5cm}
    \caption{\label{diag-hera}  
	Diagrams for $e^+ +p \to e^+ jet +X$ HERA events
     in  the $R$-parity violating SUSY model.}
  \end{center}
\end{figure}
\begin{figure}[thb]
  \vspace*{-1.5cm}
  \begin{center}
    \leavevmode
    \epsfxsize=10cm
    \epsffile{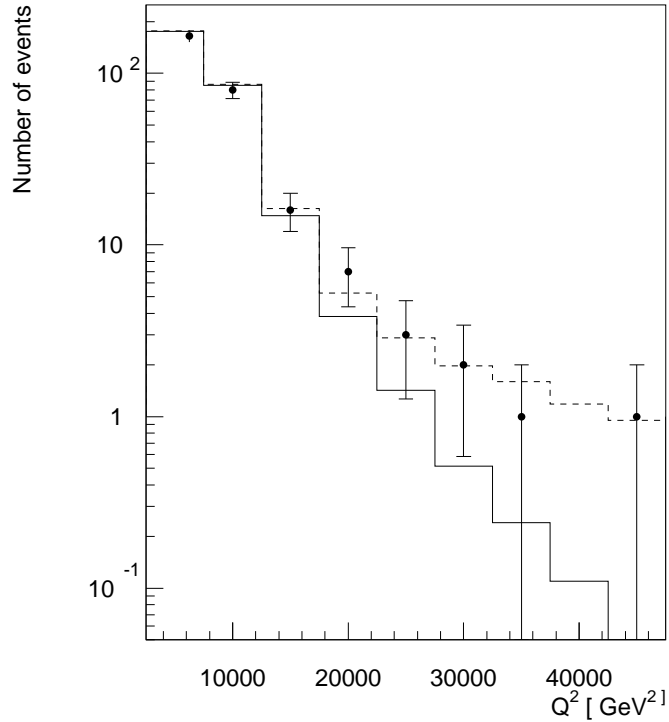}
  \vspace*{-1.0cm}
     \caption{\label{dist-q}  
	$Q^2$ distribution, predicted by SM (solid line),
     SM+$R$-parity breaking squark resonance (dashed line)
     [$\l ' =0.047$] and  data (solid dots).}
\end{center}
\end{figure}
\clearpage

\section{ $R$-parity violation at HERA}

To describe HERA anomalous events, we used data obtained by these
collaborations when 10 events survived after the $Q^2>20000$ GeV$^2$
cut, while only 4 events were expected according to Standard Model
predictions. We also took into account the integrated luminosity
ZEUS+H1 equal to 34.4~pb$^{-1}$ and an average detector efficiency 0.8.

In our calculations we used the CTEQ3M structure function~\cite{cteq}
and $Q^2$ scale $(p_e - p'_e)^2$. The total cross section describing
the observed data turned out to be 0.36~pb, while the Standard Model
gives only 0.16~pb.  Five events observed by the H1 collaboration
($Q^2>20000$~GeV$^2$) have a clear peak for  $m_{ej}=200$ GeV. ZEUS
events are more spread over $m_{ej}$, but also have a mean value around
200 GeV. That is why the natural mass of sparticles describing these
events is 200 GeV.

We have calculated the complete set of Feynman diagrams of the process
$e^+ + p \to e^+ + jet + X$ in the framework of the $R$-broken
supersymmetric model for HERA energy. The relevant graphs are shown in
fig.\ref{diag-hera}. As is naturally expected the main contribution
comes from diagrams with a resonant squark in the $s$-channel (99\%).
Contributions from the $t$-channel diagrams  are negligible (1\%) for
the case under consideration.  However, these diagrams give the same
order contribution as  a {\it nonresonant} $s$-channel  diagram with
a heavy sparticle, but this is not the subject of our study.

We use values for $R$-parity violating Yukawa couplings, which satisfy
existing limits, see table~\ref{limits}. We assumed all couplings
non-zero. Couplings  $\l'_{111}$, $\l'_{131}$, $\l'_{132}$ and
$\l'_{133}$ were fixed at their upper bounds coming from the electron
neutrino mass limits. The rest $\lambda'$s were put equal to each other
and varied in the allowed range. The presence of non-zero couplings
implies the branching ratio $BR(\tilde q \to e^+ q)$ to be smaller than
unity, even if the $R$-parity conserving decay channel is suppressed. For
the values of $\l'$'s under consideration we have $BR(\tilde t \to e^+
q) \simeq 2/7$ and $BR(\tilde c \to e^+ q) \simeq 1/3$. Due to the
smallness of $\l'_{111}$ the diagram with the $\tilde u$ exchange is
neglected hereafter.

We obtained the  value of $\l'$ equal  to 0.047 which describes the
event excess in HERA data with the 200 GeV $s$-channel squark resonance.
This value is higher than the 0.024 limit coming from~\cite{e21}. But
the limit on coupling in~\cite{e21} was obtained under the assumption
that only one $\lambda$ was non-zero, which is not the case of  this
study, and as it was shown in~\cite{ellis} this limit can be relaxed.

Figure~\ref{dist-q} shows the $Q^2$ distribution for the Standard Model
and SUSY model with broken $R$-parity tuned to describe data, and
combined ZEUS+H1 data distribution itself.

One can see that the shape of deviation from SM is perfectly
described by the $s$-channel squark resonance.

\section{ Implication for LEP200 and TEVATRON}

\begin{figure}[thb]
  \vspace*{-4.0cm}
 \hspace*{1.0cm}
  \begin{center}
    \leavevmode
    \epsfxsize=15cm
    \epsffile{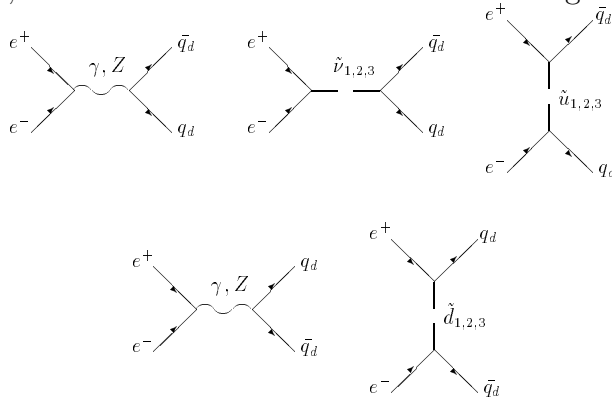}
    \vspace*{-11.5cm}
    \caption{\label{diag-lep} 
	Diagrams for the $e^+ e^-\to q\bar{q}$ process
     in the $R$-parity violated SUSY model}
  \end{center}
\end{figure}

Having in mind that processes with $R$-parity violation could occur in
$e^+ e^-$ collisions at LEP, we consider the possibility of detecting
it in the pair $b\overline{b}$ production via sparticle exchange.
Comparing to the Standard Model, some excess in dijet production is
expected. For this purpose, we have calculated the complete set of
relevant diagrams presented in fig.\ref{diag-lep}. Hereafter, we assume
that sleptons have  mass 200 GeV if not mentioned specially.

\begin{figure}[thb]
  \vspace*{-1.0cm}
  \begin{center}
    \leavevmode
    \epsfxsize=8cm
    \epsffile{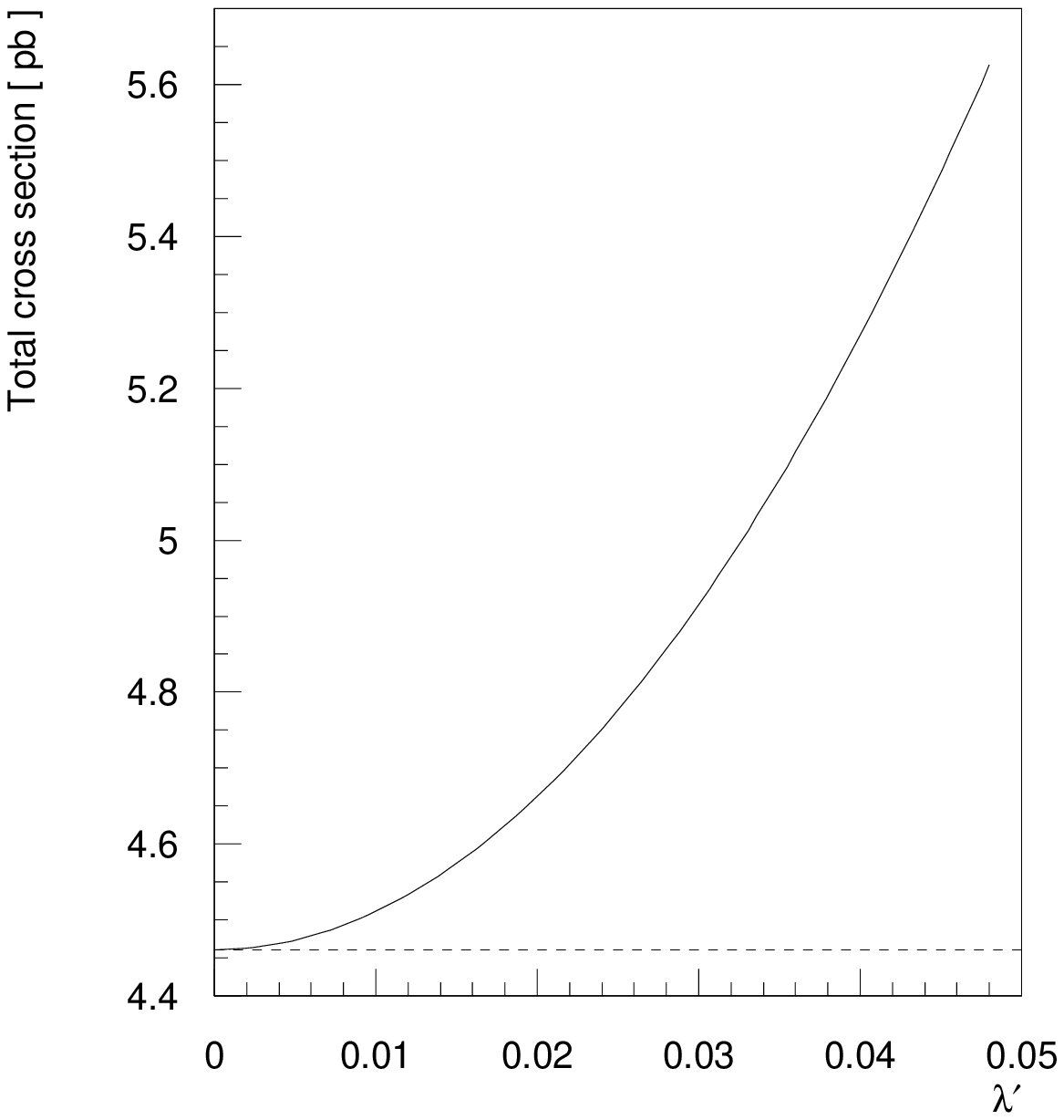}
    \put(-160,220){a)}
    \put(-140,220){$e^+ e^- \to b\bar{b}$}
    \put(-140,210){$\sqrt{s}=180$~GeV}
    \put(  70,220){b)}
    \put(  90,220){$e^+ e^- \to c\bar{c}$}
    \put(  90,210){$\sqrt{s}=180$~GeV}
    \epsfxsize=8cm
    \epsffile{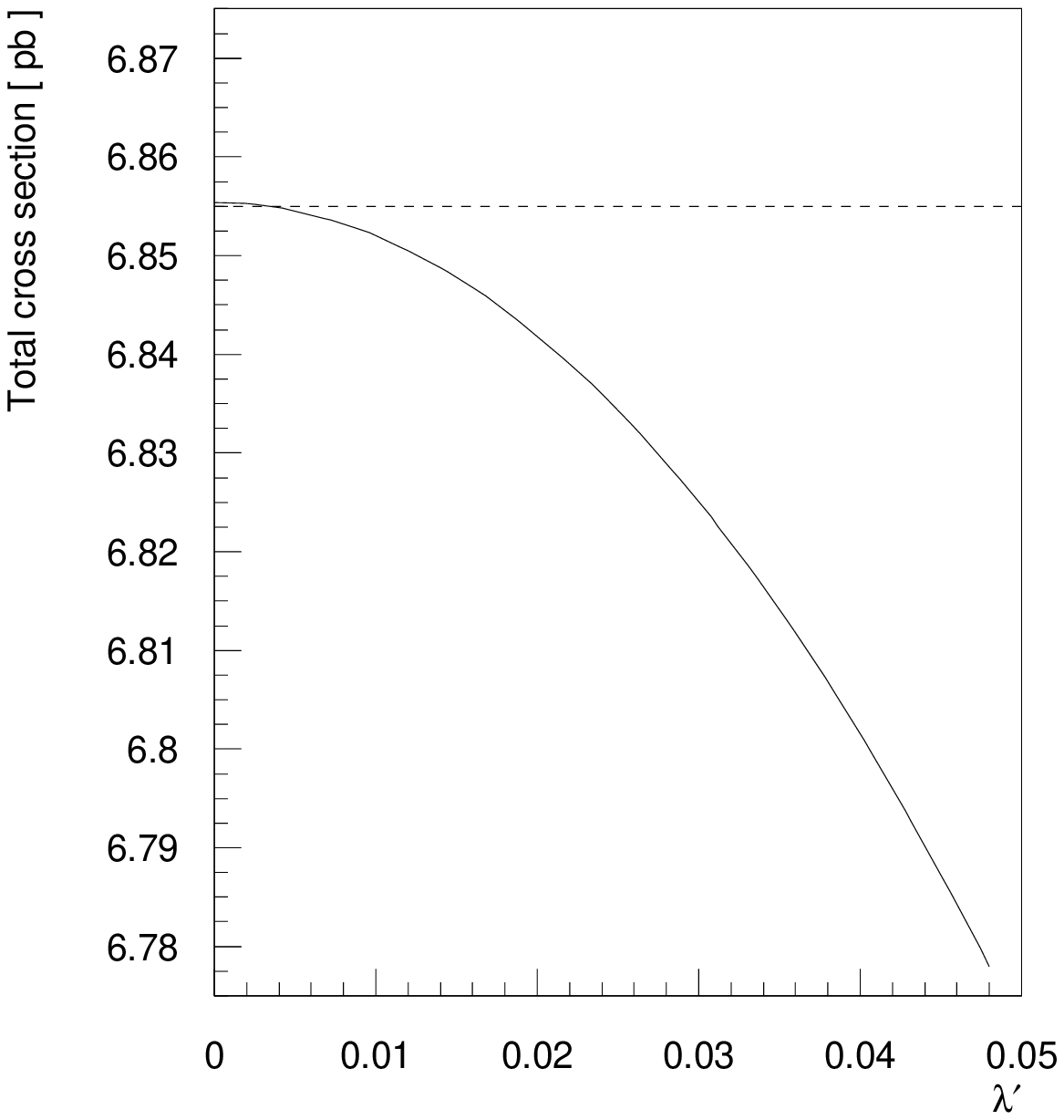}
    \vspace*{-1cm}
    \caption{\label{lep} 
	Total cross section of $e^+ e^-\to q\bar{q}$
    versus $\l'$ for down (a) and up (b) quark pair production
    at LEP. Dashed line --- SM prediction.}
  \end{center}
\end{figure}

There is a big difference in the up and down quark production. There is
an additional $s$-channel diagram with sneutrino  for the down quark
production, which makes the total cross section dependent not only on
$\l'$ but also on  $\l$ values. If we put $\l$ equal to  its upper
limit (0.2), see table~\ref{limits}, then we will have $25\%$ deviation
$\simeq 1.1$~pb of the total cross section from the Standard Model  for
$\sqrt{s}=180$~GeV LEP. In our calculation we put the value of $\l'$ to
those describing  HERA data --- 0.047. The dependence of the total $
q_d \bar{q}_d$ cross section on $\l '$  is shown in fig.\ref{lep}a). The
value of $\l$ of an order of 0.1-0.2 leads to the 15-25\% excess in the
$b\bar{b}$ production cross section. This deviation from the SM $R_b$
prediction can be measured at LEP.  As for the up quark production,
$t$-channel diagrams bring a very small negative contribution  for
values of $\l$'s  under consideration.   The dependence of the total $
q_u \bar{q}_u$ cross section as a function of  $\l'$ is shown in
fig.\ref{lep}b). One could expect some contribution of a 200 GeV SUSY
particle to the very accurately measured $R_b$ value for LEP90 which is
equal to $0.2178\pm0.0011$ and still has 1.5$\sigma$ deviation from
theoretical predictions. But it turned out that for $\sqrt{s}=90$~GeV
the contribution from additional diagrams to the $b\bar{b}$ production
was only 0.021 pb, while the total production $b\bar{b}$ rate was 9040
pb.  Thus, this additional  SUSY contribution changes $R_b$ only in the
6-th digit.


Though limits on $\l$'s are not so strict, it seems unnatural if their
values are much higher than $\l'$.  So we would like to look at  the
case when  $\l$ is small or equal to 0 and  the excess of the $b\bar{b}$
production is absent.
 For example, for $\l\simeq\l'=0.047$ the $b\bar{b}$ production
deviates from the Standard Model only by 5\%.  By a lucky
chance we have at present  three different  colliders  with energy of
an order of 1 TeV:  lepton-lepton --- LEP, lepton-hadron --- HERA and
hadron-hadron --- TEVATRON. These machines are highly complementary to
each other. We show below that possible failure to find $R$-parity
broken SUSY (if it really exists) at LEP leads to the luck at
TEVATRON in this search and vice versa.

 For TEVATRON one of the the most interesting effects of revealing
$R$-parity violation can be the resonance single top production via a
slepton decay.

\begin{figure}[thb]
  \vspace*{-3.5cm}
 \hspace*{1.0cm}
  \begin{center}
    \leavevmode
    \epsfxsize=15cm
    \epsffile{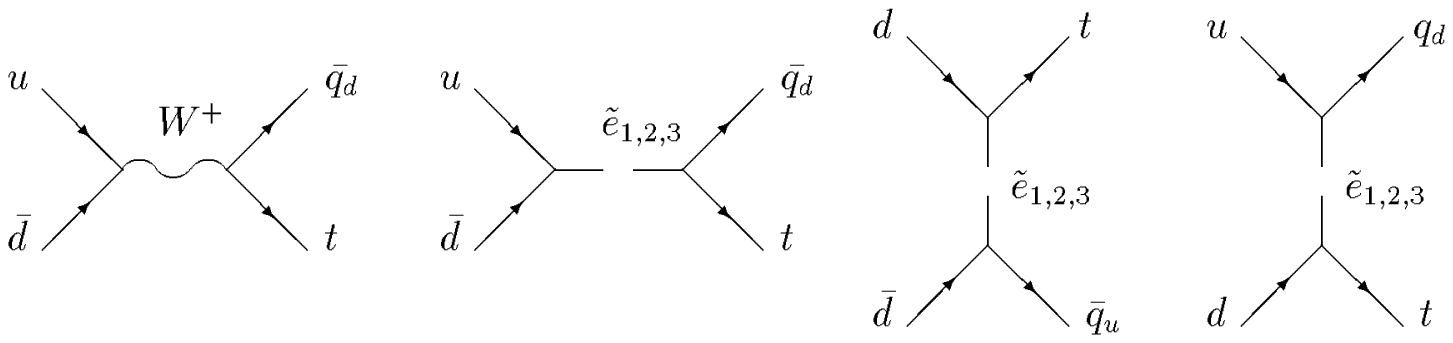}
    \vspace*{-15.0cm}
    \caption{\label{diag-tev} 
	Diagrams for $2\to2$ single top production
     processes at TEVATRON in the $R$-parity violated SUSY model.}
  \end{center}
\end{figure}

The complete set of Feynman diagrams for the $2 \to 2$ process of
single top production is shown in fig.\ref{diag-tev}.
%
%
\begin{figure}[thb]
  \vspace*{-1.0cm}
  \begin{center}
    \leavevmode
    \epsfxsize=8cm
    \epsffile{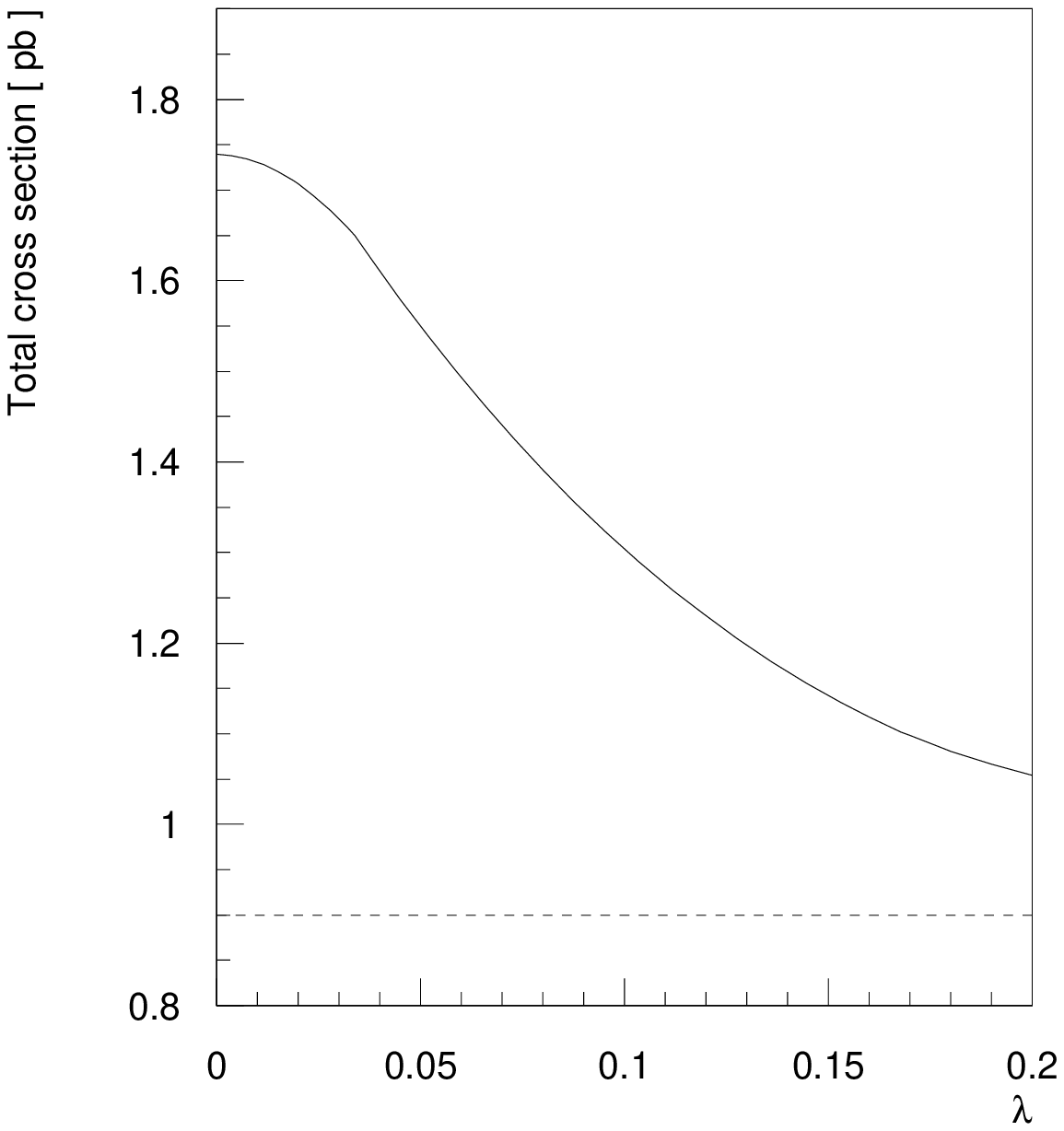}
    \put(-160,220){a)}
    \put(-140,220){$q \bar{q}' \to t\bar{q}_d$}
    \put(-140,210){$\sqrt{s}=1.8$~TeV}
    \put(  70,220){b)}
    \put(  90,220){$e^+ e^- \to b\bar{b}$}
    \put(  90,210){$\sqrt{s}=180$~GeV}
    \epsfxsize=8cm
    \epsffile{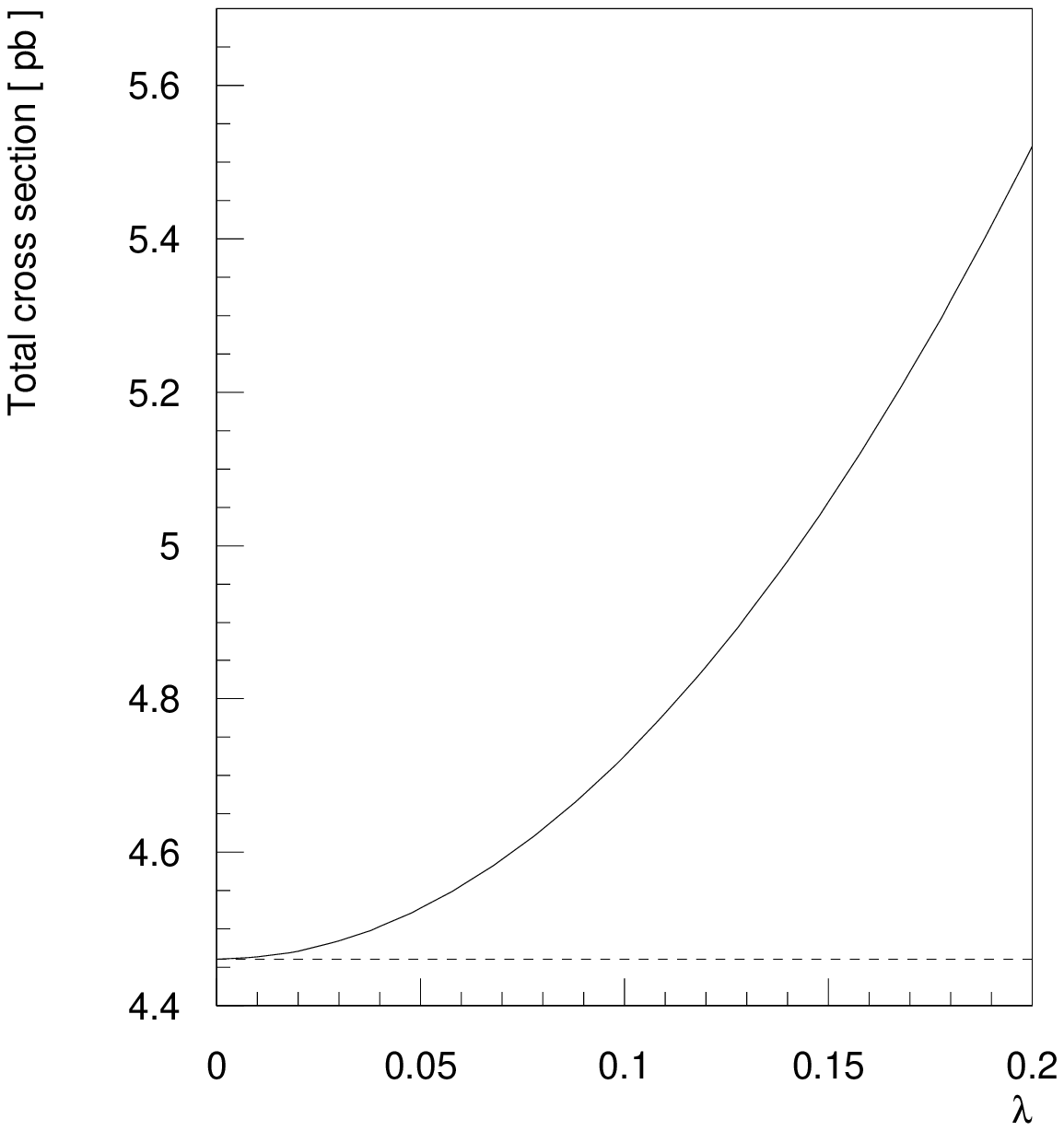}
  \vspace*{-1.0cm}
    \caption{\label{tev} 
	Total cross section of the $p\bar{p}\to t\bar{q}_d(\bar{t}q_d)$
	single top production process versus $\l$ (a) and total cross
	section of the $b\bar{b}$ production at LEP180 versus $\l$ (b).
	Dashed line shows the SM prediction.}
\end{center}
\end{figure}

%
%

In spite of that the cross section is directly proportional only to
$\l'$, there is also dependence of the cross section on the $\l$
coupling which is directly related to the width and branching ratios of
the selectron decay.  For comparatively big  $\l$ (0.1-0.2) the
selectron mainly decays leptonically and the single top production is
suppressed. In this case, the deviation in the $b\bar{b}$ production
could be measured at LEP. On the other hand, if  $\l$ is small or even
equal to 0, there is a significant contribution to the single top
production via the resonant selectron diagram. This contribution  can
be as large as 100\% of the total single top production rate of SM
$q\bar q' \to t\bar{q}_d$.  It is clearly illustrated in fig~\ref{tev}
where the dependence of this deviation versus $\l$ ($\l'$ is fixed to
0.047) is shown. If we assume the total cross section of the single top
$t\bar{b}$ $(+\bar{t}b)$ production including NLO corrections
\cite{will} (we use NLO K-factor 1.5) to be about 0.9 pb,  for
$\l$= 0.0 -- 0.1 the deviation from the SM prediction is 50--100\%. The
same effect is expected for the $W-qluon$ fusion single top production
process which is a separate subject of study and not discussed here. We
can see that by lucky coincidence LEP and TEVATRON magically complement
each other in search for supersymmetry with broken $R$-parity, which
naturally describes high $Q^2$ HERA events.

\section*{ Conclusions}

We have demonstrated that HERA events can be explained within the
Supersymmetric Standard Model with broken $R$-parity. The Yukawa
couplings responsible for processes under study do not exceed their
experimental bounds. Processes with $R$-parity breaking can also take
place in $e^+ e^-$ collisions at LEP200 and TEVATRON, which are
complementary to each other. If  $R$-parity violation really takes
place,  as has been shown above, it can be  revealed at either
TEVATRON or LEP in the near future.

\vspace{1cm}

We would like to thank D.I.Kazakov and W. de Boer for valuable
discussions and stimulation of this study. Our work has been supported
by Russian Foundation for Basic Research, grants \# 96-02-17379,
\#96-02-19773-a  and ICFPM  1996 grant.

\end{document}